\documentclass[aps,prb,a4paper,reprint,superscriptaddress]{revtex4-2}

\usepackage{amsmath,amssymb}
\usepackage{color}
\usepackage{graphicx}
\usepackage{url}
\begin{document}
\author{S. Vahid Hosseini}
\affiliation{Superconductivity Research Laboratory (SRL), Department of Physics, University of Tehran, 
North Kargar Ave., P. O. Box 14395-547, Tehran, Iran}
\affiliation{LCP-A2MC, Universit\'e de Lorraine, 1 Bd Arago, F-57078 Metz Cedex 3, France}
\author{Andrei Postnikov}
\affiliation{LCP-A2MC, Universit\'e de Lorraine, 1 Bd Arago, F-57078 Metz Cedex 3, France}
\email{andrei.postnikov@univ-lorraine.fr}
\author{Mohammad Reza Mohammadizadeh}
\affiliation{Superconductivity Research Laboratory (SRL), Department of Physics, University of Tehran, 
North Kargar Ave., P. O. Box 14395-547, Tehran, Iran}

\title{First-Principles Study of Hydrogen Dynamics in Monoclinic TiO}


\begin{abstract}
The existence of intrinsic vacancies in cubic (monoclinic) TiO suggests opportunity for
hydrogen absorption, which was addressed in recent experiments.
In the present work, based on first principle calculations, 
the preferences are studied for the hydrogen absorption sites and
diffusion paths between them. The oxygen vacancies are found to be
the primary hydrogen traps with absorption energy of $-2.87\,$eV. The plausible channels
for hydrogen diffusion between adjacent vacancy sites (ordered in the monoclinic TiO structure)
are compared with the help of calculations using the nudge elastic band 
method. Several competitive channels are identified, with barrier heights
varying from 2.87 to 3.71~eV, that is high enough to ensure relative stability
of trapped hydrogen atoms at oxygen vacancy sites.
Moreover, the possibility of adsorption of molecular hydrogen was tested and found improbable,
in the sense that the H$_2$ molecules penetrating the TiO crystal
are easily dissociated (and released atoms tend to proceed towards oxygen vacancy sites).
These results suggest that hydrogen may persist in oxygen vacancy sites 
up to high enough temperatures.
\end{abstract}

\maketitle

\section{Introduction}
\label{sec:intro}
The present work aims at elucidating hydrogen trapping and mobility
in titanium monoxide. This subject falls within a relatively small intersection of vast
domains of research. The hydrogen absorption, storage and processing in materials
is a tremendous and rapidly evolving field; a recent review 
by Abe \textit{et al.} \cite{IntJHydrEnergy44-15072} 
may serve a nice introduction into the subject. 
Transition metal oxides are omnipresent in materials science through their different manifestations,
over which a concise review by Goodenough \cite{ChemMater26-820}
offers an efficient guideline.
In what regards specifically the hydrogen problematics,
transition metal oxides typically appear as elements of surface protection of metals
or as catalytic agents, e.g., in a promising research field 
of water splitting on oxide-based electrocatalysis \cite{JMaterChemA8-13415},
explained, e.g., in recent reviews by Zhu \textit{et al.} \cite{EneEnvironSci13-3361},
Song \textit{et al.} \cite{ChemSocRev49-2196} and Shang \textit{et al.} \cite{SustainEneFuels4-3211}.
Burke \textit{et al.} \cite{ChemMater27-7549} summarized the trends 
in the activity of different transition metal oxides as catalysts 
for the \textit{oxygen evolution reaction}.
The phenomenon of \textit{hydrogen spillover} came into discussion, in which
molecular hydrogen dissociates at the metal surface, and protons diffuse into
the catalytic oxide support -- see, e.g., Ref.~\citenum{JAmChemSoc143-9105}
concerning the process on MgO, or Ref.~\citenum{JPhysChemLett10-7285} for the reaction on VO$_2$.
Whittingham \cite{SolidStateionics168-255} described hydrogen motion in metal oxides;
Nolan and Browne \cite{CurrentOpinionElectrochem21-55} overviewed chemical reactions 
relevant in oxides in relation with hydrogen energy problematics.

Still narrowing the subject, the works from this category dealing with titanium  
are not numerous, and predominantly concern the dioxide.
Oelerich \textit{et al.} \cite{JAlloysComp315-237,AdvEnginMater3-487} studied catalytic activity
of $3d$ metal oxides, including TiO$_2$, and concluded that a tiny addition 
of them noticeably enhanced hydrogen sorption kinetics of nanocrystalline magnesium.
Yin \textit{et al.} \cite{ChemPhysChem9-253} studied, in experiment supported by
first-principles calculations, the adsorption of hydrogen 
on the surface of TiO$_2$ with possible diffusion into the depth. 
Feng \textit{et al.} \cite{ACS_Catal8-4288} looked into effect of oxygen vacancies in TiO$_2$
on the enhancing of electrocatalytic activity for \textit{hydrogen evolution reaction} (HER).
Oxygen vacancies also play an important role in shaping the 
\textit{oxygen evolution reaction} essential for water splitting. A recent review
on these issues by Zhu \textit{et al.} \cite{NanoEnergy73-104761} covers,
among a long list of oxides studied, works on titanium dioxide and oxide-nitride.
Hu \textit{et al.} \cite{ChemMater19-1388}
praised micro- and mesoporous Ti oxides as ``in many ways ideal candidates 
for hydrogen storage because they can be made from inexpensive and light metal Ti, 
and the surface area, pore size, and wall thickness can be systematically controlled''.
Li \textit{et al.}\cite{NatCommun10-3149} studied the hydrogen evolution activity
in epitaxially grown Ti$_2$O$_3$ polymorphs. 

Titanium monoxide enters the picture as Swaminathan \textit{et al.}\cite{ACS_Catal6-2222}
reported that strongly reduced titania (with nominal composition TiO$_{1.23}$, the crystal
structure of which reveals X-ray diffraction peaks of disordered TiO) exhibits enhanced
HER activity.
Recently, Skripov \textit{et al.} \cite{JAlloysComp887-161353} 
investigated hydrogen absorption in substoichiometric TiO 
(TiO$_{0.72}$H$_{0.30}$ and TiO$_{0.96}$H$_{0.14}$)
by a combined use of X-ray and neutron diffraction, neutron vibrational spectroscopy 
and nuclear magnetic resonance. This work led to the following conclusions:
$(i)$ in both nearly stoichiometric and strongly
substoichiometric (oxygen-deficient) samples, hydrogen atoms resided exclusively at O vacancy sites;
$(ii)$ the hydrogenation of nearly stoichiometric, originally B1 disordered, phase 
provokes an emergence of an ordered (Ti$_5$O$_5$ monoclinic) phase
coexisting with the disordered (B1-structure) TiO;
$(iii)$ hydrogen diffusion seems to be insignificant;
$(iv)$ vibrations of hydrogen atoms occur throughout a broad range
of frequencies, presumably as a manifestation of different environments
(a possible symmetry-lowering off-center displacement of the trapped
hydrogen; a completeness or not of its coordinating Ti octahedron) 
occurring at different O vacancy sites.

TiO can be crystallized in cubic B1 (NaCl-type) structure, allowing broad limits of deviation 
from stoichiometry in the nominal formula TiO$_x$, namely (citing 
Watanabe \textit{et al.} \cite{ActaCryst23-307})
$0.9\,{\leq}\,x\,{\leq}\,1.25$ at 990$^{\circ}$C and $0.7\,{\leq}\,x\,{\leq}\,1.25$ at 1400$^{\circ}$C.
As mentioned further in Ref.~\citenum{ActaCryst23-307}, even in equiatomic compound 
about 15\% of both titanium and oxygen sites are vacant. The vacancies are randomly distributed 
at high temperatures, however on rapid cooling from 1400 to 990~$^{\circ}$C, they get ordered, 
resulting in a monoclinic phase, which is in fact the underlying B1 with 1/6 of sites 
on each sublattice being vacant.
While deficiency on either cation or anion sublattice is not uncommon in materials,
the coexistence of vacancies in both sublattices is relatively rare and,
among the metal monoxides, is known to occur only in TiO \cite{JPhysChemSolids25-1397,PSSB224-R1},
VO \cite{PRB5-2775} and NbO \cite{JETPLett111-176}.

Theoretical studies of TiO have a long history.
Neckel \textit{et al.} \cite{JPhysC9-579} performed self-consistent electronic structure
calculation of TiO (among other transition metal oxides and nitrides with B1 structure,
assumed perfect and complete). This revealed the essential in the placement and composition
of different energy bands.
Leung \textit{et al.} \cite{PRB54-7857} compared the relative stability under pressure
of several Ti- and O-deficient (stoichiometric) ordered phases, including the ``true'' 
monoclinic Ti$_5$O$_5$ and several more symmetric ordered-vacancies phases;
the energy preference of the monoclinic phase has been demonstrated.
Andersson \textit{et al.} \cite{PRB71-144101} offered a more detailed analysis of
different vacancy-ordered phases, in comparison with a model of disordered alloy
(of vacancies) on each sublattice. 
(One can add in this relation 
that a much earlier attempt of a ``vacancy-alloying''
approach to electronic structure of TiO, non self-consistent and within 
virtual crystal approximation, was undertaken by Schoen and Denker\protect\cite{PhysRev184-864}).
Graciani \textit{et al.} \cite{PRB72-054117} in an almost simultaneous work reiterated
the established reasons for the structural preference of the monoclinic phase.
Kostenko \textit{et al.} \cite{JSolidStateChem204-146} concentrated on comparing the ordered 
monoclinic Ti$_5$O$_5$ phase against the B1 vacancy-disordered one, whereby 
the latter was simulated by averaging the results over 20 different supercells
of (40$\times$Ti+40$\times$O) atoms (plus 8$\times$Ti+8$\times$O vacancy sites).
Under this perspective, we considered the monoclinic TiO phase to be a perfect
model system -- not too simplistic yet well defined -- for studying 
hydrogen trapping, vibration, and diffusion. 
A different choice, that of disordered B1 lattice, would make the results
too much dependent on the particular model(s) of disorder used.
An ambition of the present work is to challenge the conclusions formulated
by Skripov \textit{et al.} \cite{JAlloysComp887-161353}, on the basis of first-principles
calculations, and/or to give these qualitative conclusions a numerical expression.
Specifically, we probe adsorption energies of hydrogen at different sites,
calculate vibration modes of the crystal doped with hydrogen,
identify plausible diffusion paths and estimate corresponding energy barriers.

In its concept, our work has certain similarities with \textit{ab initio}
simulation by Kajita \textit{et al.} \cite{JChemPhys127-104709} 
done for hydrogen adsorption at the TiO$_2$ surface,
in the sense that different trapping sites have been identified, and the energy barriers
between them explored. The difference is in the crystal structure of the underlying oxide
and in that we considered 3-dimensional crystal in our simulation, and not a surface
represented by a slab.

The present work is organized as follows. 
Section \ref{sec:struc} explains the crystal structure, adsorption sites and the paths
between them.
Section \ref{sec:comput} outlines technical details of first-principles calculations. 
Section \ref{sec:result} presents the results of hydrogen binding energy and 
intersite barriers, with its impact on diffusion. 
The conclusions are summarized in section \ref{sec:conclu}.

\section{Crystal Structure and Geometry of Vacancies}
\label{sec:struc}

\begin{figure*}[t!]
\centerline{\includegraphics[width=0.94\textwidth]{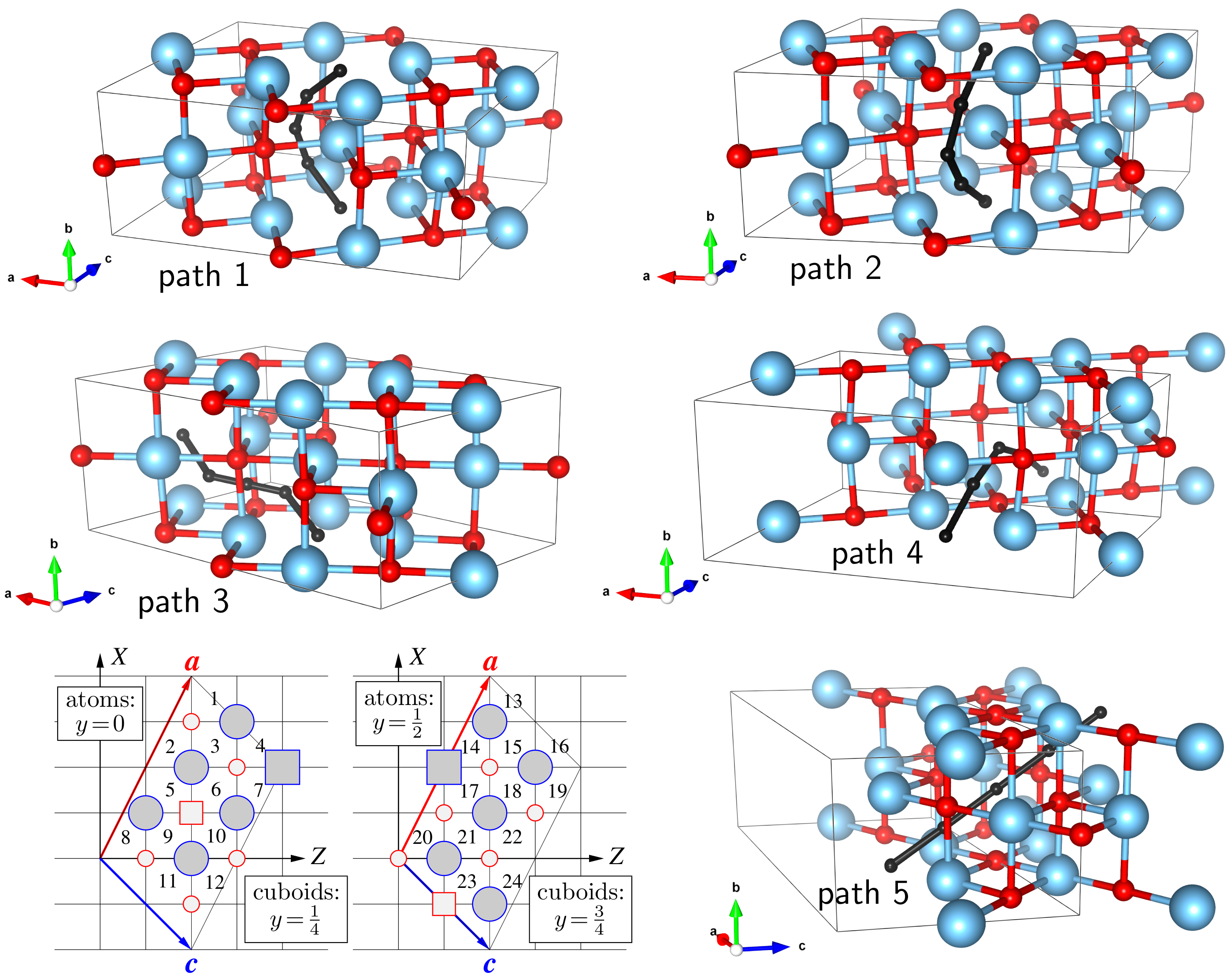}}
\smallskip
\caption{\label{fig:path}
Crystal structure of monoclinic TiO, in side views (Ti atoms shown large blue 
and O atoms small red), with five different paths (shown in black color as connected 
intermediate positions) between adjacent oxygen vacancy sites,
and (two panels in the bottom left) as two consecutive $(010)$ planes, with atoms 
shown as circles and vacancy sites shown as squares. 
The cuboids (centered $y\!=\!\tfrac{1}{4}$ and $y\!=\!\tfrac{3}{4}$)
between consecutive atomic layers are numbered, for reference in the text.}
\end{figure*}

The stable monoclinic structure of nominally stoichiometric TiO
(which holds however in the concentration range TiO$_{0.7}$ through TiO$_{1.25}$)
has been identified by Watanabe \textit{et al.} \cite{Nature210-934,ActaCryst23-307}.
In terms of the lattice parameter of the nominal underlying cubic (B1) lattice, 
and in the setting to be used in the following, the monoclinic lattice comes about 
spanned by $\mathbf{a}\!=\![2\,0\,1]$, $\mathbf{b}\!=\![0\,1\,0]$, 
and $\mathbf{c}\!=\![\bar{1}\,0\,1]$ 
vectors. Correspondingly, the unit cell contains 
$3{\times}4=12$ of both anionic and cationic sites, of which only 10 (of each species) are
occupied. Further on, $a/b\!=\!\sqrt{5}$, $c/b\!=\!\sqrt{2}$, and 
$\beta=\pi\!-\!\arctan(3)=108^{\circ}26'$. These relations, as we will see,
are slightly modified due to the presence of vacancies.
The distribution of vacancies centers the $(a,b)$ face 
(space group $C2/m$, unique axis $b$), hence the primitive cell
in fact includes five Ti and five O atoms. Their arrangement over Wyckoff positions
with coordinates refined in experiment can be found in Ref.~\citenum{ActaCryst23-307}.
The unit cell, in several side views (with different diffusion paths, see below)
and as a $[010]$ projection of two consecutive atomic planes $y\!=\!0$ and $y\!=\!\tfrac{1}{2}$,
is shown in Fig.~\ref{fig:path}. 
The choice of the unit cell is such that
Ti vacancy is at the origin. These projection figures specify,
for further reference, the numbering of Ti$_4$O$_4$ cuboids
(eventually with vacancies, represented by squares instead of circles, at some vertices) 
centered in the $y\!=\!\tfrac{1}{4}$ and $y\!=\!\tfrac{3}{4}$ planes.

A priori, the possible hydrogen absorption sites are expected to possess certain symmetry;
plausible candidates are interstitial positions (inside the Ti$_4$O$_4$ cuboids, probably with 
one or two corner atoms missing), the Ti vacancy sites, or the O vacancy sites. All these possibilities
have been explored (see the details below), with the conclusion that
the oxygen vacancies are the ground-state configurations, Ti vacancies are
metastable local minima, and the other trial configurations end up squeezed out 
into one or the other of the two mentioned. Anticipating the question of hydrogen diffusion between
the oxygen vacancy sites, we can already elaborate on the possible trajectories
connecting such adjacent positions. Fig.~\ref{fig:path} summarizes all possible paths
connecting adjacent O vacancies, shown as straight-line fragments through the centers of 
different (numbered) cuboids. In practical calculations
using the Nudged Elastic Band (NEB, see below) technique, 
the trajectories will be shortened / smoothened.

Making reference to numbered cuboids (at the bottom left of Fig.~\ref{fig:path}) which ``pave''
two alternating (010) planes of the monoclinic structure, the paths under discussion
can be identified as follows.

Both path 1 and path 2 connect the O vacancy sites separated by $[0\,1\,0]$ move,
circumventing the Ti atom in between. The path 1 goes (``upwards'', from the O vacancy position
with $y\!=\!0$) through cuboids 6 and 18 (or, equivalently, 9 and 21); the barrier 
is obviously anticipated on squeezing through the Ti$_2$O$_2$ face in between. The path 2 
goes through cuboids 5 and 17 (or, equivalently, 10 and 22). The saddle point at $y\!=\!\tfrac{1}{2}$
occurs now on squeezing through a face which is not complete, because the cuboids in question
share a titanium vacancy. Consequently, the path is expected to be pushed either towards
or away from  this vacancy, and the barrier to become lower than for path 1.

A ``watershed'' between the paths 1 and 2, with their corresponding saddle points,
would be on the Ti--O bond, breaking which is likely to cost much energy.

Path 3 is a $[\tfrac{1}{2}\,\tfrac{1}{2}\,0]$ diagonal move, through the cuboids 6 and 3
(or, equivalently, 9 and 11); the bottleneck is squeezing through a complete Ti$_2$O$_2$ face
underway. Consequently, the barrier is expected to be similar to that on path 1.

Path 4 goes in general $[0\,0\,1]$ direction, making a bow around the O atom situated
half-way, through the cuboids 10 and 5 (or, equivalently, through the cuboids 22 and 17
under the $y\!=\!0$ plane). This path is likely to go through (or, rather, just close to)
the Ti vacancy site (at $y\!=\!\tfrac{1}{2}$) shared by the two cuboids in question.
In this sense, the path 4 resembles path 2; the difference is that an oxygen and not 
titanium atom is circumvented.

Path 5 goes straight along $[\,011]$ through e.g. the cuboids 10 and 17,
via the Ti vacancy site situated half-way (in the corner shared by the both cuboids). 
We note in this relation that passing \textit{exactly} through the Ti vacancy on this path
is imposed by symmetry (that was not the case of path 2 and 4), therefore the barrier height 
must match the difference of hydrogen absorption energies in the Ti and the O vacancies. 
We'll come back to this point in discussing the results below.  

In practical calculation of hydrogen absorption energies and diffusion,
we were careful to use sufficiently large supercells so that the hydrogen atom
could be treated as an isolated impurity, and the effects of spurious
periodicity within the hydrogen ``sublattice'' be minimized.

\section{Computational Details}
\label{sec:comput}
Electronic total energies and forces (for conjugated-gradient structure relaxation and
the extraction of force constants) were calculated within the density functional theory (DFT),
using the generalized gradient approximation (GGA) for exchange-correlation functional 
in the flavor of Perdew--Burke--Ernzerhof (PBE) \cite{PRL77-3865}.

Whereas in many calculations done for TiO$_2$ and Ti$_2$O$_3$,
the inclusion of correlation effects beyond the conventional GGA was considered
necessary and was treated within ``GGA+$U$'' scheme or using hybrid functionals
(see, e.g., Hu and Metiu\cite{JPhysChemC115-5841}
for a review), no such need seem to apply to metallic titanium monoxide.

A large part of our calculations have been done with
the help of the Quantum ESPRESSO (QE) package \cite{JPCM21-395502,JPCM29-465901},
using plane waves as basis functions,
and otherwise, so that some results would have been cross-checked,
with the {\sc Siesta} 
code \cite{JPCM14-2745,JChemPhys152-204108}, which relies on atom-centered localized 
basis functions. The ultrasoft pseudopotentials \cite{PRB41-7892} were employed 
in QE calculations, whereby Ti $3s$, $3p$, $3d$, $4s$, O $2s$, $2p$ and H $1s$ states 
were included as valence ones. In {\sc Siesta} calculations, norm-conserving 
pseudopotentials \cite{PRB43-1993} were used, with the attribution of valence states 
as indicated above, except for the Ti$\,3s$ states, which were treated as core. 
The idea of using two methods 
evolved naturally from a need to apply adequate tools to different tasks,
but eventually offered an opportunity to compare parallel predictions as well, 
which turned out to be helpful in assessment of credibility of numerical results. 
In fact, even if two codes operate at the same level of theory (DFT/GGA),
details of technical implementation may be responsible for slightly
different numerical results. Whereas some parameters ($\mathbf{k}$-mesh,
plane-wave cutoffs) can be systematically enhanced to yield neccessary precision,
the others remain somehow an experience-guided free choice. Such are the generation
of pseudopotentials, not identical in two methods, and a construction of basis functions
in {\sc Siesta}. Whereas QE is robust and free from ambiguity in what regards the choice
of bases, {\sc Siesta} with its compact yet efficient basis sets might be interesting
for trading its applicability to very large systems for an affordable
loss in accuracy. In our present compromise, we place on QE our expectations of, a priori,
higher accuracy, recognizing at the same time that the size of systems treated with QE
risks to be not sufficiently large for making a good model case of isolated impurity.
If the methods used are, in a sense, complementary, their combination may help to reasonably estimate 
the error margin due to inherent technical differences, and to judge about
the general credibility of results. 

In QE calculations
we used $1{\times}2{\times}2$ supercells (40 TiO formula units), and in {\sc Siesta}
calculations -- $1{\times}3{\times}2$ ones (60 TiO formula units),
which helped to ensure a more even separation between impurities on the lattice.
The $\mathbf{k}$-mesh used for summations over the Brillouin zone 
according to the Monkhorst--Pack scheme \cite{PRB13-5188} was more dense in the
second case ($6{\times}5{\times}5$ divisions along the reciprocal lattice vectors)
than in the first one ($4{\times}4{\times}4$ mesh points). The planewave cutoff 
for the basis set construction in QE was set to 60~Ry; the cutoff for kinetic energy and 
charge density expansion to 720~Ry. The {\tt MeshCutoff} parameter for the expansions 
of residual charge density in {\sc Siesta} was set to 300~Ry. The convergence for energy 
was chosen as $10^{-5}$~eV between two ionic steps, 
and the maximum force allowed on each atoms is 0.01~eV/{\AA}. Comparing with earlier works, 
we can state that our QE calculation setup is closest to that used by 
Kostenko \textit{et al.} \cite{JSolidStateChem204-146} in what regards the code applied,
the exchange-correlation potential and the supercell size. The differences are
that there was no extrinsic impurities considered in Ref.~\citenum{JSolidStateChem204-146},
nor optimization done for lattice parameters (of supercells simulating disorder).
Therefore we can address the reader to Fig.~7b of Ref.~\citenum{JSolidStateChem204-146} 
for inspecting the local densities of states (DOS) in the ordered phase,
which exhibit a characteristic pseudogap at the Fermi level, discussed in some
of the works cited. Partial DOS in the presence of hydrogen impurity, obtained
with {\sc Siesta}, will be shown and discussed below. 
Incorporation energy of a hydrogen atom at a vacancy site is expressed as
\begin{equation}
E^I_{\rm H}=E_{\rm TiO+H}-E_{\rm TiO}-E_{\rm H}\,,
\label{eq:01}
\end{equation}
where $E_{\rm TiO+H}$ is the total energy of TiO supercell containing a
H atom, $E_{\rm TiO}$ is the total energy of pristine TiO supercell, 
and $E_{\rm H}$ is the energy of isolated H atom. 
Applying such formula to total energy results obtained
with {\sc Siesta} (or, with any other method employing atom-centered basis functions)
demands to correct for basis set superposition error (BSSE, see Ref.~\citenum{ChemRev94-1873}).
In practical terms of our case, this involved placing hydrogen basis sets
(``ghost atoms'' not carrying core charges nor electrons)
at two relevant (trial) positions for hydrogen adsorption.
Either one (for calculating $E_{\rm TiO+H}$) or none (for $E_{\rm TiO}$)
of these positions were in fact occupied by hydrogen atom, the rest being ghosts.
Correspondingly, $E_{\rm H}$ stems from a calculation for the same
supercell of the same shape, in which a single (spin-polarized) genuine hydrogen atom
cohabits with 121 ghosts. 

The inspectrion of minimum energy paths (MEP) 
connecting distinct local minima (ty\-pi\-cal\-ly over a saddle point) can be, 
from the side of theory, conveniently done by the
Nudged Elastic Band (NEB) method, implemented in the QE code. 
Ref.~\citenum{JChemPhys128-134106} offers an overview of NEB and other related schemes,
which deal with a sequence of intermediate ``images'',
e.g., conformations of the system subject to interplay of forces ``along''
and ``away from'' the path, ensuring the smoothness and the shortness of the latter.
In the present calculations, we considered 13 images
and (independently) 25 images, in order to check the stability of results against
this parameter; ${\tt k\_\,min}$ and ${\tt k\_\,max}$ switches for elastic bands
were chosen at 0.2 and 0.3~Ha, respectively.

\section{Results \& Discussion}
\label{sec:result}
Our calculations included unconstrained structure relaxation
of (for reference purposes) pristine monoclinic TiO and, in appropriately enlarged 
supercell, of hydrogen impurity tentatively placed in various lattice positions.
It turned out that the hydrogen remains (the most) stable at oxygen vacancy site,
but also at a local energy minimum at titanium vacancy site.
Hydrogen escapes from other symmetric positions, e.g., in the centrum
of a Ti$_2$O$_2$ face, or cutting a Ti--O bond.

\subsection{Equilibrium crystal structures; comparison of methods}

Some results reported below (lattice relaxations, energy barriers) have been obtained
with two methods, QE and {\sc Siesta}. Even if QE, free from ambiguities 
in constructing the basis functions,  might be considered as ultimately more reliable
(for the given supercell geometry and similar pseudopotentials), we prefer to
expose the {\sc Siesta} results along with those by QE. This will help to get some idea
of the ``credibility margin'', in view of {\sc Siesta} being applied to larger-size
supercells and yielding some supplementary properties to those explored by QE.

\begin{table*}[t!]
\caption{\label{tab:cages}
Sizes of empty or H-occupied octahedral cages, according to GGA calculations.
For comparison, lattice parameters mapped onto a single monoclinic cell are given. 
See text for details.}
\medskip
\begin{tabular}{
c@{\hspace*{6mm}}
*{2}{c@{\hspace*{5mm}}c@{\hspace*{5mm}}cp{4mm}}c@{\hspace*{3mm}}c@{\hspace*{1mm}}c}
\hline 
 & \multicolumn{3}{c}{\parbox[b]{3.9cm}{%
   \begin{center}dimensions\\ (Ti--Ti in {\AA}) \\of ${\square}_{\rm O}$ along... \end{center}}} &&
   \multicolumn{3}{c}{\parbox[b]{3.9cm}{%
   \begin{center}dimensions\\ (O--O in {\AA}) \\of ${\square}_{\rm Ti}$ along... \end{center}}} && 
   \multicolumn{3}{c}{\parbox[b]{3.8cm}{%
   \begin{center}reduced lattice \\ parameters ({\AA})$^{\ast}$ \end{center}}} \\*[-10pt]
    \cline{2-4} \cline{6-8} \cline{10-12}
 System & $[10\bar{1}]$ & $[102]$ & $[010]$ && $[10\bar{1}]$ & $[102]$ & $[010]$ &&    
          $a$ & $b$ & $c$ \rule[-4pt]{0pt}{14pt} \\      
\hline
\multicolumn{12}{c}{Quantum ESPRESSO calculations ($1{\times}2{\times}2$ supercells)} \\
pristine               & 3.905   & 4.102 & 4.163 &&
                         4.247   & 4.328 & 4.163 && 9.319 & 4.163 & 5.845 \\ 
H@${\square}_{\rm O}$  & $\underline{3.945}$ & 4.098 & $\overline{4.110}$ &&
       4.241   & 4.316 & 4.162  && 9.330 & 4.162 & 5.839 \\
H@${\square}_{\rm Ti}$ & 3.889   & 4.091 & 4.171 && $\underline{4.337}$ & \underline{4.404} &
$\underline{4.309}$   && 9.329   & 4.165 & 5.853 \\*[2pt] 
\hline
\multicolumn{12}{c}{{\sc Siesta} calculations ($1{\times}3{\times}2$ supercells)} \\
pristine               & 3.926   & 4.239 & 4.226 &&
                         4.340   & 4.489 & 4.247 && 9.507 & 4.224 & 5.954 \\ 
H@${\square}_{\rm O}$  & $\underline{3.997}$ & 4.242 & $\overline{4.185}$ &&
       4.321   & 4.479 & 4.252  && 9.510 & 4.229 & 5.958 \\
H@${\square}_{\rm Ti}$ & 3.921   & 4.241 & 4.230 && $\underline{4.398}$ & \underline{4.535} &
$\underline{4.362}$   && 9.511   & 4.224 & 5.953 \\*[2pt] 
\hline
\multicolumn{12}{c}{X-ray diffraction by Watanabe \textit{et al.}$^{\dagger}$} \\
pristine    & 4.026    & 4.197   & 4.142 &&
                         4.233   & 4.256 & 4.142 && 9.340 & 4.142 & 5.855 \\ 
\hline
\end{tabular}
\\*[2mm]
$^{\ast}$Crystallographic angle was within $+0.06\%$ in QE calculations and within $-0.5\%$
in {\sc Siesta} calculations from the experimental value $\beta=107^{\circ}32'$ as reported in
$^{\dagger}$Ref.~\citenum{ActaCryst23-307};
the original attribution of crystallographic parameters is changed in the table to match that
in later works.
\end{table*}

Table~\ref{tab:cages} shows the structure parameters of nominal (pristine) TiO, 
in comparison with experiment,
and the (predicted) modification of structure under insertion of hydrogen. 
{\sc Siesta} tends to overestimate the lattice parameters (by at most 2\% against the results
by Watanabe \textit{et al.}\cite{ActaCryst23-307}; by $\simeq\,5\%$
for the volume) whereas QE seems to perform much better, within 1\% of experiment values
for every lattice parameter and the volume. However, the dimensions of cages around
oxygen vacancies are estimated with {\sc Siesta} at least not worse than with QE 
(judging again by comparison with experiment).
What seems important for the following is that the distortion of vacancy cages
(measured between two opposite atoms flanking either the vacancy, or the hydrogen atom 
occupying the vacancy site) upon insertion of hydrogen follows the same pattern 
according to either QE or {\sc Siesta} and seem reliable beyond the calculation ``noise'':
a roughly uniform expansion (by 1--2\%) occurs around hydrogen occupying the Ti vacancy,
whereas around hydrogen at the O vacancy site, an expansion along the long diagonal 
of the unit cell (i.e., parallel to $[10\bar{1}]$) is combined with compression
in the perpendicular direction ($[102]$). In Table~\ref{tab:cages},
those distances which noticeably (by $\simeq\,1\%$) shrink as compared to situation 
in pristine lattice are overlined in the table; those which expanded 
(by at least the same margin) are underlined. The average lattice parameters 
(shown in the three last columns of Table~\ref{tab:cages}
and the dimensions of the hydrogen-free vacancy cages remain practically unchanged.

\subsection{Adsorption energies, lattice relaxation, electronic structure}
\label{subsec:energies}
Hydrogen incorporation energies calculated according to Eq.~(\ref{eq:01}) yield, 
with total energies from QE calculations, $-2.87\,$eV at the O vacancy site (hence
energetically favorable insertion) and $+0.75\,$eV at the Ti vacancy site
(hence costing energy). Even if straightforwardly identifiable from the point
of view of calculation, these values might be not so easy to relate to
experiment, because of ambiguity and difficult reproducibility of
reference situations, e.g., molecular hydrogen penetrating
the crystal, dissociating, etc.
``Straightforward'' {\sc Siesta} calculations, performed without correcting
for BSSE, are in qualitative agreement with QE results, yielding
$-3.13\,$eV and $+0.20\,$eV for adsorption at the O vacancy and
the Ti vacancy sites, respectively. {\sc Siesta} calculations staged to 
minimize the systematic error by excluding the BSSE, e.g., done with the equal number
and positions of (either real or ghost) atoms in the three situations
relevant for Eq.~(\ref{eq:01}), result in adsorption energies
of $-2.86\,$eV (H@${\square}_{\rm O}$) and $+0.73\,$eV (H@${\square}_{\rm Ti}$). 
Such good agreement between results of a planewave-basis and a localized-basis
methods, even if ideally anticipated, is not always secured technically, 
and can serve here as an additional argument for the credibility of results.

\begin{figure}[b!]
\centerline{\includegraphics[width=0.50\textwidth]{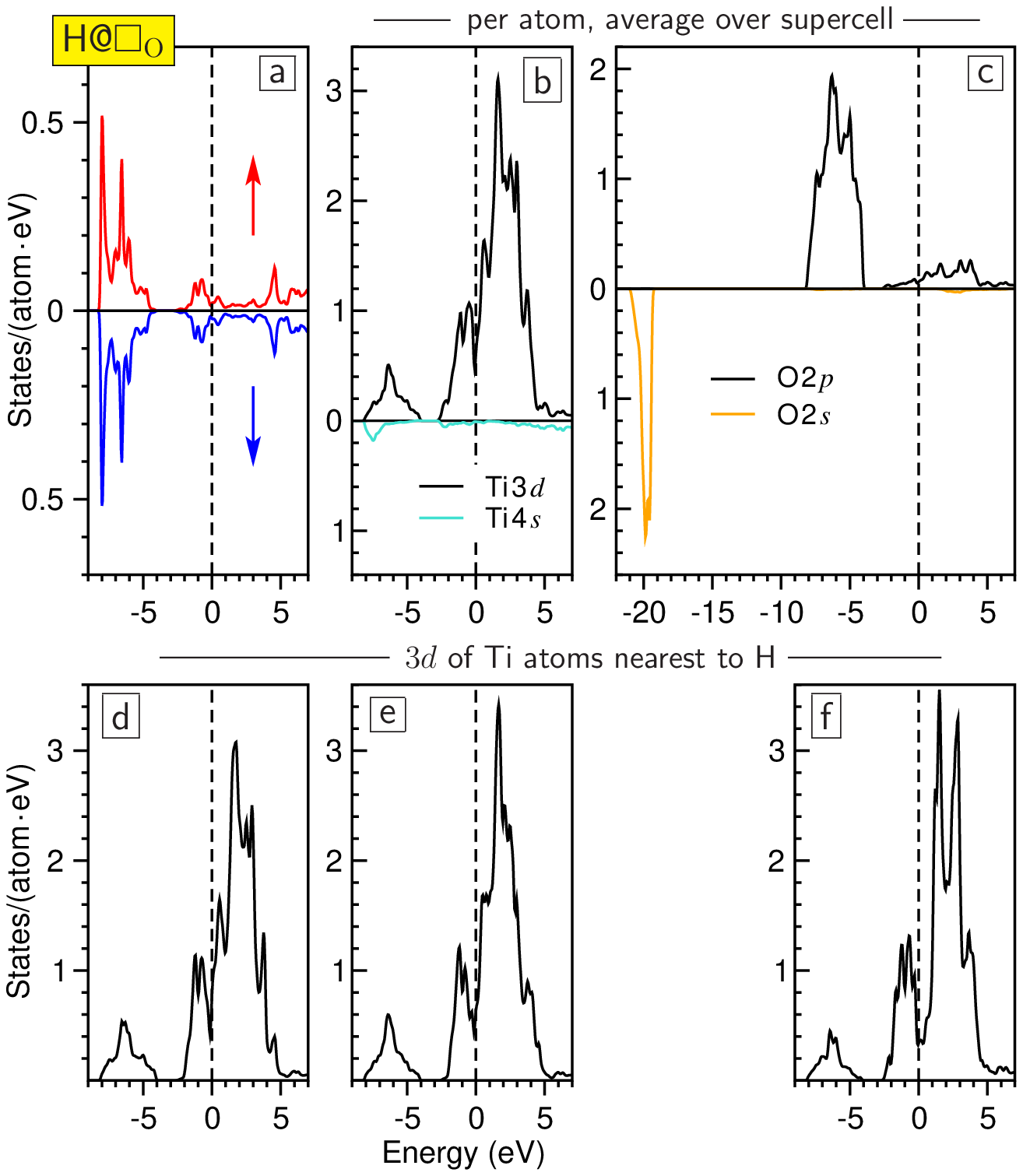}}
\smallskip
\caption{\label{fig:DOS_O}
Partial densities of states (DOS) in relaxed $1{\times}3{\times}2$ (120 atoms) supercell of monoclinic TiO,
containing also a hydrogen atom at the oxygen vacancy site, from a {\sc Siesta} calculation.
Zero energy (dashed vertical line) corresponds to the Fermi energy.
The upper row includes (a) spin-resolved DOS of H atom (in fact non-magnetic in this position);
(b) $3d$ and $4s$ DOS of Ti (averaged over all Ti sites in the supercell); 
(c) $2s$ and $2p$ DOS of O (averaged over all O sites in the supercell). The bottom row
depicts the $3d$ DOS for three symmetrically distinct pairs of Ti atoms 
octahedrally bordering the vacancy site
occupied by hydrogen, namely (d) atoms situated at ${\pm}\tfrac{1}{6}(\mathbf{a}\!-\!\mathbf{c})$,
(e) -- at ${\pm}\tfrac{1}{6}(\mathbf{a}\!+\!2\mathbf{c})$, (f) -- at ${\pm}\tfrac{1}{2}\mathbf{b}$,
in units of nominal translation vectors of the monoclinic structure (cf. Fig.~\ref{fig:path}). }
\end{figure}

The variations of total energies have their origin in the fine details
of the electronic structure, even if the direct relation might be difficult to trace. 
Fig.~\ref{fig:DOS_O} depicts local densities of states at the hydrogen atom at the O vacancy,
its nearest Ti neighbors, and both these species averaged over the supercell.

An inspection of Mulliken populations 
done with the {\sc Siesta} code (comparing the fully relaxed situation
of a hydrogen insertion with the pristine crystal and ghost atom at the vacancy site)
shows an inflow of about 0.1 electrons onto H at the O vacancy site from its
surrounding Ti atoms. 
Interestingly, certain analogy can be found between this system and the rocksalt TiH hydride,
addressed in a number of first-principles simulations:\cite{JLessComMet88-269,PRB66-144107}
the placement of H-related band at $\simeq\,6-8\,{\rm eV}$ below the Fermi level,
the charge transfer towards H, the general shape and energy placement of the Ti$3d$ band,
the order of magnitude of the formation energy.
Especially Smithson \textit{et al.}\cite{PRB66-144107}
paid attention to elucidating different contributions to the hydride
formation energy, including conversion of the metal structure to fcc,
expansion to the optimal lattice parameter, and chemical bonding to hydrogen.
Under this angle, a H-occupied O vacancy in titanium monoxide is already in
``favorable'' fcc (rocksalt)-like enviroment of Ti atoms, whereby the distances
between the latter (cf. Ti--Ti size of the O vacancy cage in Table~\ref{tab:cages})
roughly matches the lattice parameter of the rocksalt titanium hydride,
(4.10~{\AA}, according to Table~II of Ref.~\citenum{PRB66-144107}).
A discussion around Fig.~6 in the same work specifies the details of charge loss
by Ti $e_g$ orbitals in favor of H-centered spherical distribution. Much of this
discourse applies to our system as well. Focusing on the details specific for 
H at the O vacancy in TiO, we note that the 
Ti DOS is characterized by a dip at the position of the Fermi level,
marked in earlier calculations of monoclinic TiO\cite{PRB49-16141,PRB54-7857}.
Among the six Ti atoms neighboring the inserted hydrogen, the two closest to it
(compacted towards hydrogen along $[010]$, see Table~\ref{tab:cages})
develop the most pronounced difference (two distinct peaks just above the Fermi level,
see Fig.~\ref{fig:DOS_O}f)
from the pristine (or, lattice-averaged, cf. Fig.~\ref{fig:DOS_O}b) Ti DOS.

\begin{figure}[t!]
\centerline{\includegraphics[width=0.50\textwidth]{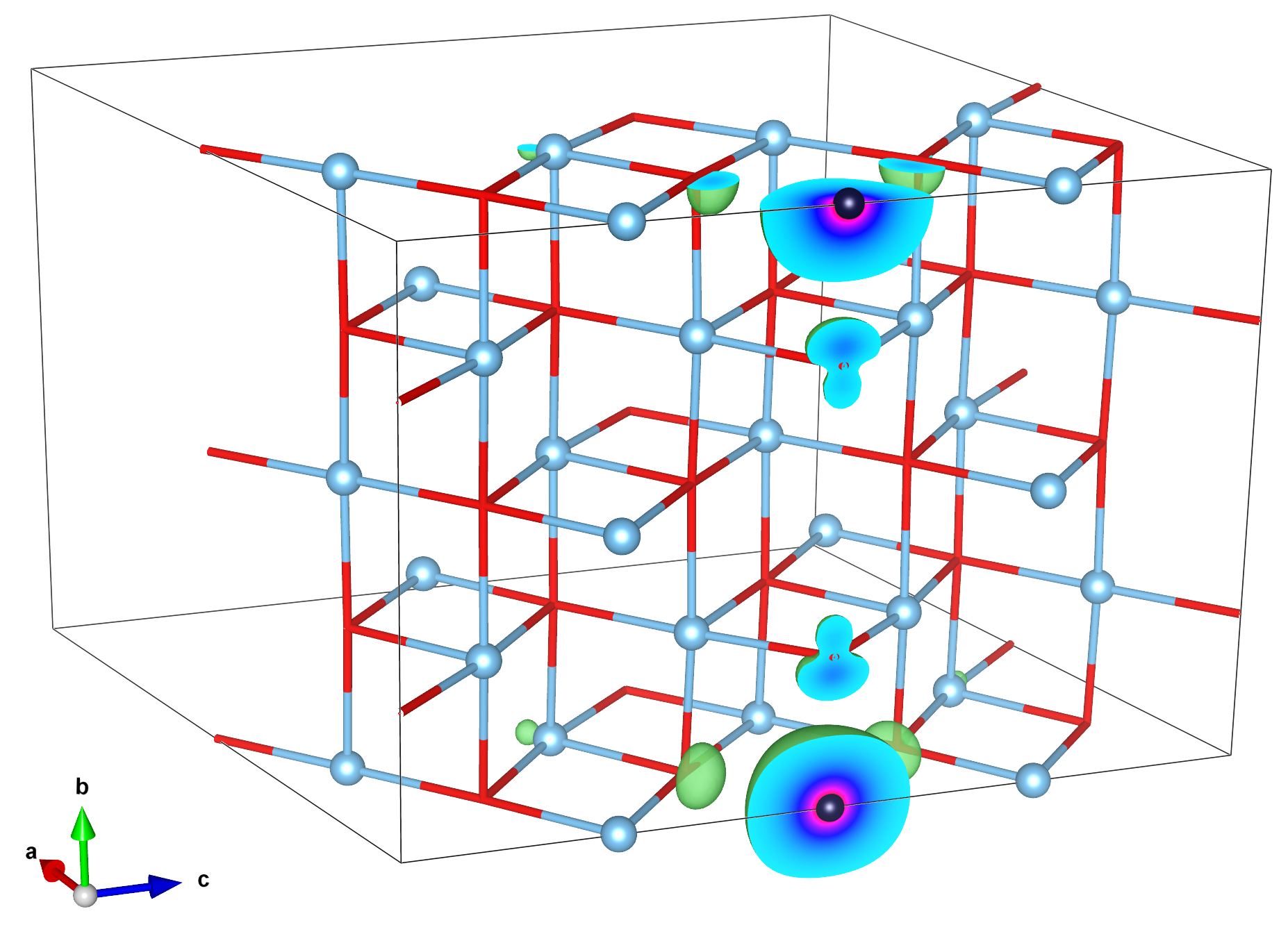}}
\smallskip
\caption{\label{fig:SpinDen}
Spin density map calculated with QE for the supercell with hydrogen (small black sphere)
at the Ti vacancy site. Ti atoms are marked by blue spheres; O atoms by red crosses in the wireframe. 
The isosurface shown in green corresponds to the level of $0.003\,\mu_{\rm B}\mbox{\AA}^{-3}$; the color schema
in the cuts by the unit cell faces means enhancement from cyan to magenta.
As the isolevel increases, the surfaces rapidly become spherical and compact around the H site.
In total, there is $1\,\mu_{\rm B}$ per H atom.}
\end{figure}

\begin{figure}[t!]
\centerline{\includegraphics[width=0.50\textwidth]{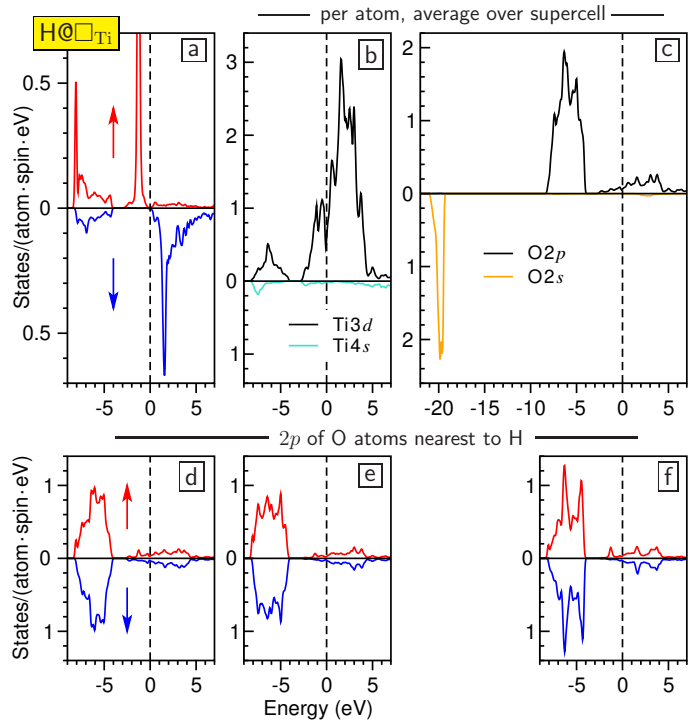}}
\smallskip
\caption{\label{fig:DOS_Ti}
Similar to Fig.~\ref{fig:DOS_O}, for the case of a hydrogen atom at the titanium vacancy site.
The system possesses the magnetic moment of $1\;{\mu}_{\rm B}$. The bottom row depicts 
$2p$ DOS of O atoms around the Ti vacancy site, namely (d) atoms situated 
at ${\pm}\tfrac{1}{6}(\mathbf{a}\!-\!\mathbf{c})$,
(e) -- at ${\pm}\tfrac{1}{6}(\mathbf{a}\!+\!2\mathbf{c})$, 
(f) -- at ${\pm}\tfrac{1}{2}\mathbf{b}$.
}
\end{figure}

The other possible placement of H in TiO, at the Ti vacancy site, is a metastable one,
with elevated total energy yet corresponding to a local energy minimum. The peculiarity
of this configuration is the local magnetic moment of $1\,{\mu}_{\rm B}$ inherited from
a free hydrogen atom, due to a relative absence of chemical bonding and of charge transfer 
in either direction between H and its surrounding oxygen atoms.
We'll see in the following that the stability of the magnetic solution 
is rather sensitive to an off-center displacement of hydrogen. Fig.~\ref{fig:SpinDen} 
shows the spatial distribution of the spin density which is
strongly localized on the impurity, yet slightly spills out onto the neighboring 
O atoms, especially onto those at $\pm\tfrac{1}{2}\mathbf{b}$. This observation is further reinforced 
by an inspection of partial DOS concerning H at the Ti site and its neighbors (Fig.~\ref{fig:DOS_Ti}).
The most remarkable feature of the magnetic structure is a narrow majority-spin state 
just below (by ${\simeq}1\,{\rm eV}$) the Fermi level. This state, strongly localized
at hydrogen, also manifests itself in the partial DOS of O neighbors, primarily 
those along $[010]$ -- cf. Fig.~\ref{fig:DOS_Ti}(f). The spatially resolved 
density of electronic states, integrated in energy just over this peak (done with {\sc Siesta},
not shown here) closely reproduces the full spin density revealed by Fig.~\ref{fig:SpinDen}.

Despite the practical absence of the charge transfer, the electronic shells of oxygen atoms
are spin polarized to (together over the six atoms) $0.12\,\mu_{\rm B}$. 
This polarization is two times stronger on the ``out-of-plane'' atoms $[\,$situated at
${\pm}\tfrac{1}{2}\mathbf{b}$, marked (f)$\,]$ in Fig.~\ref{fig:DOS_Ti} than on the
``in-plane'' atoms, marked (d) and (e), as is also revealed by the manufestation of 
the features below the Fermi energy in the corresponding partial DOS.

\begin{figure}[t!]
\centerline{\includegraphics[width=0.45\textwidth]{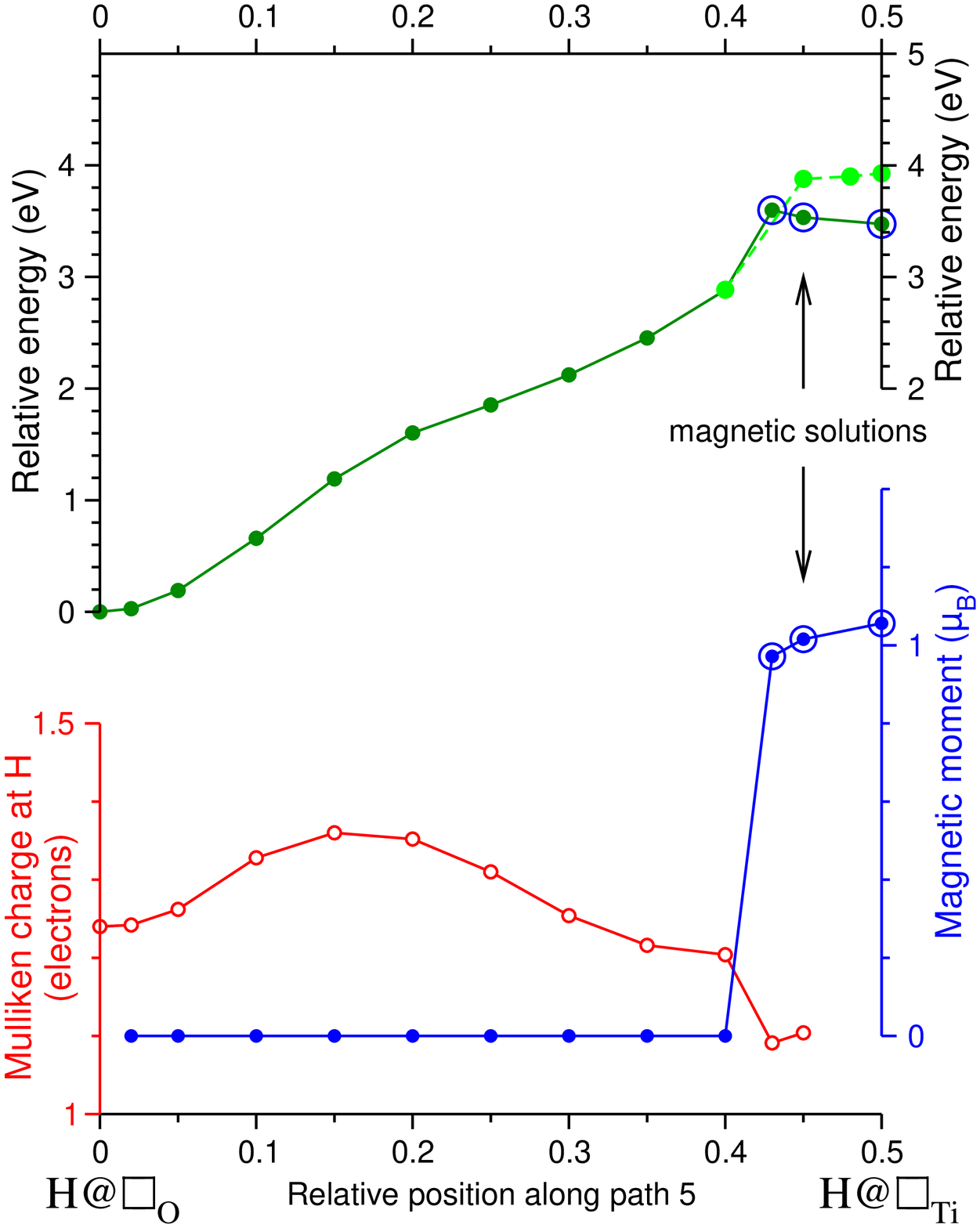}}
\smallskip
\caption{\label{fig:path5}
Variation of total energy, magnetic moment and Mulliken charge at the H atom 
in the course of the latter's continuous displacement from O-vacancy to Ti-vacancy position
(after {\sc Siesta} calculations, allowing the relaxation of atoms along the path).
Magnetic solutions, which survive only in small vicinity of the Ti vacancy, are marked
by circles. The energy values for their non-magnetic counterparts are indicated by light green dots.
}
\end{figure} 

For a more systematic insight into the variance of properties at two possible sites 
for hydrogen, we show in Fig.~\ref{fig:path5} a simulated smooth transition between them, 
along the path 5 of Fig.~\ref{fig:path}.
Calculated energy profile, the Mulliken charge at the H atom and the total magnetic moment
stem from a row of {\sc Siesta} calculations, in which H atom was fixed at the intermediate
positions along the path, its nearest neighbors were free to relax, and more distant atoms 
in the supercell were kept frozen. This can be considered as a forerunner of more general
and flexible NEB calculations over the ensemble of paths, to be discussed below.
One clearly sees that the relative energy increases along the path,
consistently with the energetics discussed above in subsection~\ref{subsec:energies},
until eventually the magnetic solution sets up, resulting in a modest energy lowering.
The ``potential well'' around the Ti vacancy site is $\approx\,1\,${\AA} wide
and $\simeq\,0.1\,{\rm eV}$ (or $\simeq\,1160\,{\rm K}$) deep, 
so that the zero-point energy for hydrogen in it is about 30~meV, 
hence this site seems plausible for hypothetical trapping of hydrogen.

One notes that the magnetic solution sets up abruptly and maintains
the value of almost exactly $1\,\mu_{\rm B}$, even if the system, being metallic,
does not impose an integer value of the magnetic moment. This reinforces the hypothesis
that hydrogen at Ti vacancy site behaves as a trapped free atom. Consistently with
this view, the Mulliken charge, somehow elevated throughout the ``hydride-like''
part of the path, drops down to 1.1, a nearly nominal value for a free atom.

The bump in the value of the Mulliken charge at 1/3 of the distance from
O-vacancy to Ti-vacancy occurs as the hydrogen atom squeezes between three titanium
atoms, across a face of a Ti-octahedron delimiing the O-vacancy cage.
This bottleneck is characterised by a charge flow from Ti to H; the total energy curve
goes markedly upwards from the direct linear slope. At 2/3 of the distance,
the hydrogen atom passes through a triangle of oxygen atoms without a noticeable
hybridisation or charge loss, and, being released into the O-octahedron around 
the Ti vacancy site, it looses extra charge and recovers its free-atom magnetic moment.
A slight rearrangement of states within the occupied fraction of Ti$3d$--O$2p$ bands
accounts for a lowering of total energy. Gradually uprising energies of forced zero-spin solutions
(cf. light green dots at the right edge of the upper panel in Fig.~\ref{fig:path5})
demonstrate that magnetism of H at the Ti-vacancy site is essential for the 
(meta)stability of this configuration.

\subsection{Phonons}

\begin{figure}[!b]
\centerline{\includegraphics[width=0.50\textwidth]{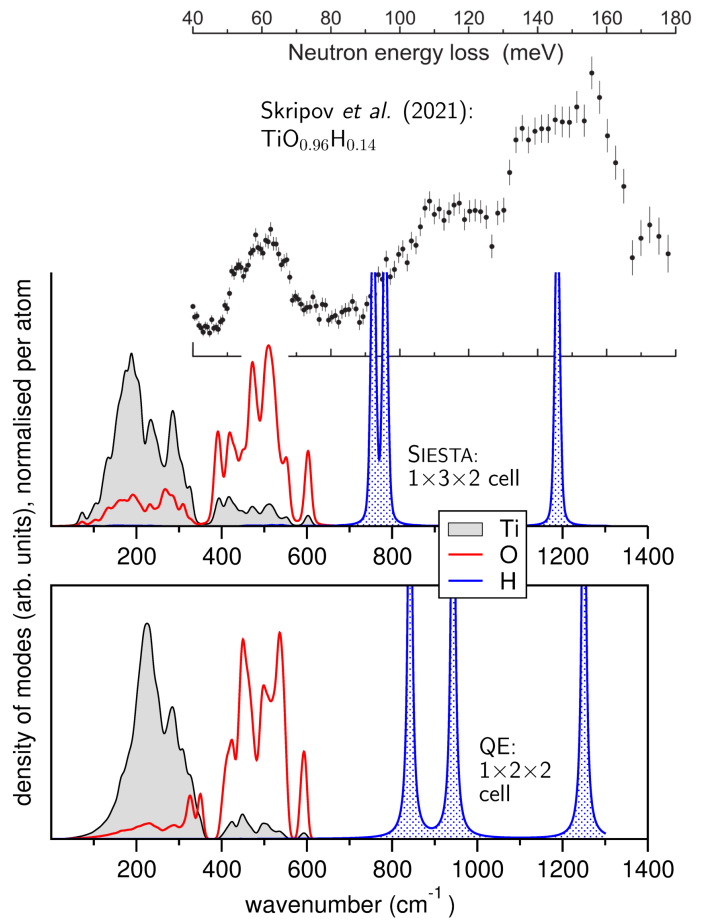}}
\smallskip
\caption{\label{fig:phonons}
Densities of vibration modes of TiO with H occupying an oxygen vacancy site
(one per supercell indicated), calculated by QE (bottom panel) and {\sc Siesta} (middle panel).
Neutron scattering spectrum of monoclinic TiO$_{0.96}$H$_{0.14}$, shown for comparison
in the upper panel, is reproduced from Fig.~5 of Ref.~\citenum{JAlloysComp887-161353} --
copyright Elsevier (2021). See text for detail.
}
\end{figure}

For additional insight, and also in order to offer some discussion of interesting  
neutron energy loss spectra published by Skripov \textit{at al.} \cite{JAlloysComp887-161353},
we calculated the density of vibration modes for H-doped (one atom per supercell)
monoclinic TiO. The QE calculation used density functional perturbation method,
and {\sc Siesta} -- frozen phonon calculation. In both these cases, just $\Gamma$ phonons 
for the corresponding supercell have been calculated, yielding 243 and 363 modes,
respectively. The resulting densities of modes are depicted
in Fig.~\ref{fig:phonons}, slightly broadened (with halfwidth parameter of 5~cm$^{-1}$)
for better visibility. The abovementioned neutron scattering spectrum is reproduced
for comparison, with energy axes properly aligned.

One notes an encouraging agreement between the Ti/O parts of the vibration spectra
obtained by two methods, the degree of such agreement being not a priori obvious
in view of certain difference in calculating the electronic structure and the assessment 
of phonons. As was correctly anticipated in Ref.~\citenum{JAlloysComp887-161353},
the low-energy (${\simeq}\,60\,$meV) peak in their neutron scattering spectra 
(cf. Fig~4 and 5 of the work cited) is due to optical vibrations of oxygen atoms.
Moreover, based on inspection of vibration patterns in different modes one can now
conclude that the split-off peak at the top of the oxygen vibration band
at ${\simeq}\,600\,{\rm cm}^{-1}$, which is also pronounced in the experimental spectrum,
reveals the vibrations of 4-coordinated oxygen atoms, i.e., those bordering  
to Ti vacancies at ${\pm}\tfrac{1}{2}\mathbf{b}$.

Hydrogen vibrations make three distinct lines, a close ``doublet'' 
at ${\simeq}\,800-900\,{\rm cm}^{-1}$ and a markedly distant peak at ${\simeq}1200\,{\rm cm}^{-1}$.
This general picture holds for both {\sc Siesta} and QE calculations, even if
precise frequencies differ (756, 784, $1188\,{\rm cm}^{-1}$ in {\sc Siesta} calculation
vs. 842, 943, $1250\,{\rm cm}^{-1}$ according to QE). 
This picture agrees well with the observation by Skripov \textit{et al.}\cite{JAlloysComp887-161353} 
that ``... the spectra look like superpositions of two broad bands centered at about 
115 meV and about 155 meV'', taking into consideration a ``volatility'' of calculated frequencies
under slightest modification of calculation conditions, due to the smallness of the proton mass.
A direct inspection of the modes in question reveals that the vibrations occur along the Ti--Ti
axes of the Ti$_6$ octahedron, i.e., $[102]$, $[010]$ and $[10\bar{1}]$, respectively.
The variance in vibration frequencies nicely correlates with the sizes of Ti$_6$ cages
(cf. Table~\ref{tab:cages})
and the hence following steepness of the potential well confining the hydrogen vibrations.
Namely, the softest of the three vibrations occurs along the direction of the largest 
Ti--Ti distance, that is $[102]$ according to {\sc Siesta} and $[010]$ according to QE.
This ``contradiction'' is not crucial because these two cage dimensions are not much
discriminated. On the contrary, the markedly split-off hardest mode represents the vibration 
along $[10\bar{1}]$, the shortest cage size, in both {\sc Siesta} and QE calculations.

Skripov \textit{et al.}\cite{JAlloysComp887-161353} 
suggest that ``... the high-energy band
at about 155 meV ... may be ascribed to the characteristic environment of H 
in the monoclinic phase'', to which statement we now offer an adjusting explanation.

It was emphasized in Ref.~\citenum{JAlloysComp887-161353} that
``... the volume of the voids formed by oxygen vacancies is too large for H atoms, 
so that a hydrogen atom can be easily displaced from the geometrical center of the vacancy''
and ``...for a considerable number of H atoms, the nearest-neighbor environment 
is no longer octahedral''. In fact we show that off-center displacement is not crucial
for explaining broad peaks and that, even when preserving an octahedral
environment, slightest modifications of the latter's shape and size due to all possible
distant imperfections in the lattice are capable to generate a broad band
of hydrogen vibration frequencies, typical for a spectra of real materials.

\subsection{Dissociation of H$_2$ molecule}

In the study of hydrogen uptake and diffusion in metals, it is generally accepted that 
the H$_2$ molecule is first physisorbed at the surface and may then overcome 
the activation barrier for dissociation (see, e.g., a discussion around Fig.~1 
of Kirchheim and Pundt \cite{PhysMetall25-2597}). Hydrogen atoms may further be chemisorbed
and eventually diffuse into the material. 
As in the discussion about hydrogen diffusion (see next subsection), quantum effects
may be important in overcoming the activation barrier; an early representative work 
to this effect, simulating dissociative adsorption (of H$_2$ on Cu surface), was done 
by Mills and J{\'o}nsson \cite{PRL72-1124}. We do not address this issue
here in details, but, as the geometry and the properties of TiO are somehow different
from common metals, we wonder whether a hydrogen molecule may fit into, and survive intact within,
the oxygen vacancy cage.

\begin{figure}[b!]
\centerline{\includegraphics[width=0.45\textwidth]{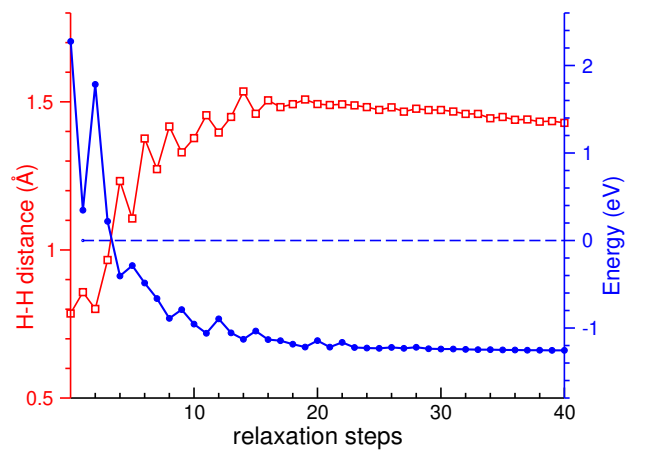}}
\smallskip
\caption{\label{fig:H2disso}
Evolution of interatomic distance (left scale, red squares) and total energy 
(right scale, blue dots) of an H$_2$ molecule placed in an oxygen vacancy,
after a {\sc Siesta} calculation.
Zero energy corresponds to a situation with 
\textcolor{blue}{a}
single H atom relaxed at oxygen vacancy site 
and another one being free. See text for details. 
}
\end{figure}

As it turns out in the course of conjugate-gradient total energy minimization
(see Fig.~\ref{fig:H2disso}, one example out of several trial ones starting from
different initial orientations), the molecule eventually ``almost dissociates'', 
doubling the nominal H--H bond length to ${\simeq}1.4\,${\AA}. 
It looks like every hydrogen atom preferentially ``couples'' to three closest Ti 
atoms, in the spirit of the remark made in the previous subsection about the oxygen
vacancy cage being too large for a hydrogen atom.
We did not manage to reproduce a definite dissociation, when one of the H atoms would flee 
the Ti$_6$ cage delimiting an oxygen vacancy site, because this would require an energy inflow 
to overcome a barrier (see next subsection); however, such event seems to be within grasp
for a molecular dynamical high-temperature simulation. Indeed, an energy step of ${\simeq}1.2\,$eV
(see Fig.~\ref{fig:H2disso}) separates the ``weakened''/expanded H$_2$ molecule from a situation 
when one of its constituent H atoms is promoted out of the crystal. An irrevocable dissociation 
and diffusion away from the original site of an atom remaining within the crystal 
may presumably cost less energy.

\subsection{Energy barriers for hydrogen atom hopping over oxygen vacancy sites}

\begin{figure*}[t!]
\centerline{\includegraphics[width=0.92\textwidth]{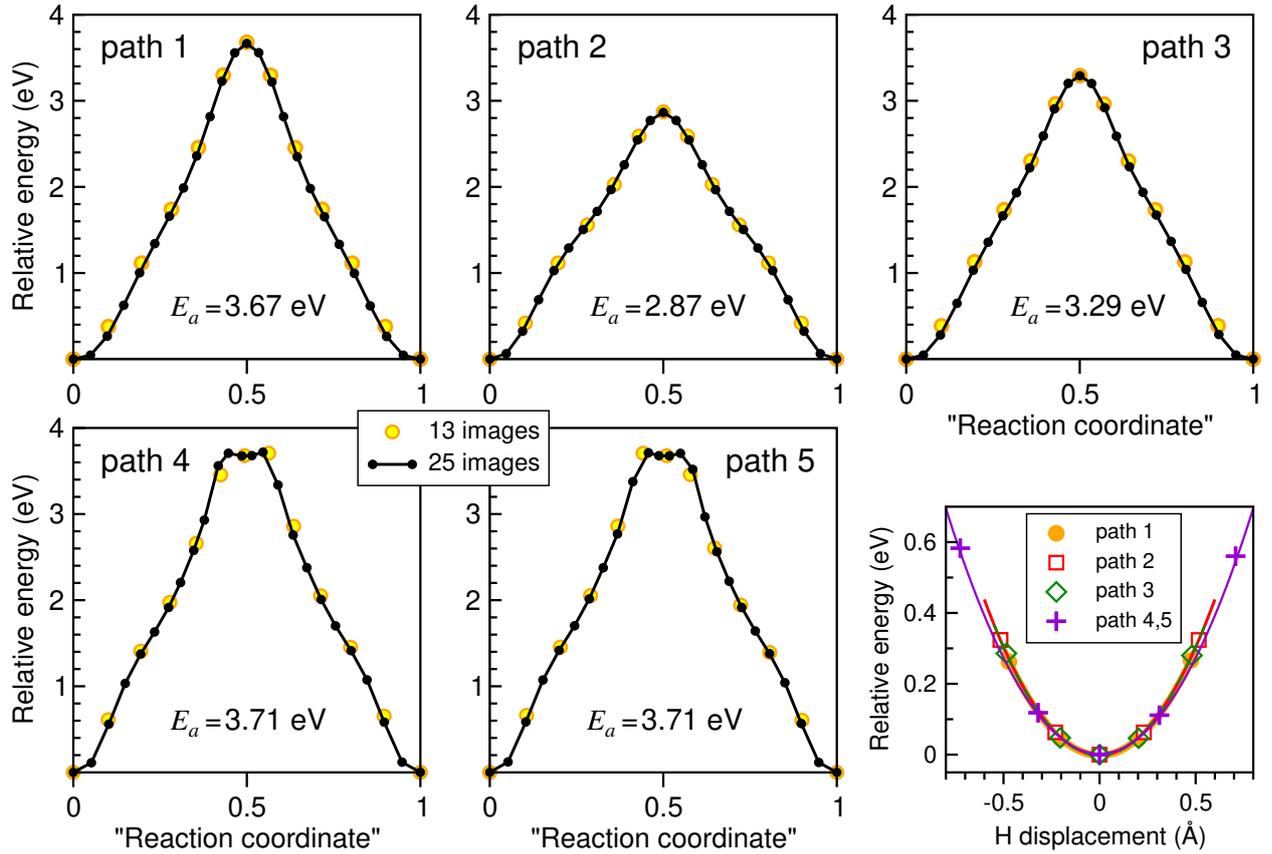}}
\smallskip
\caption{\label{fig:NEB_results}
NEB energy profiles for H atom displacement along the five paths specified
in Fig.~\ref{fig:path}. Results of calculations performed with 13 and with 25 images
along each path are shown for comparison. The barrier height is indicated for each path.  
In the bottom right panel, the total energy is traced as function of absolute displacement
of the hydrogen atom from equilibrium position, near the beginning and the end of each path. 
}
\end{figure*}

We turn now to the discussion of scenarios, outlined earlier in Section \ref{sec:struc},
of how a hydrogen atom, initially trapped in oxygen vacancy site, can be brought onto an
adjacent similar site. Practical calculations have been done according to NEB formalism implemented
in QE code; comparison of results obtained with 13 and with 25 images along the path
gives us credibility in what regards the barrier height, whereby the barrier profile obtained
with 25 images is more smooth and permits to discuss some details. All the paths must
in principle be left-right symmetric; this is not exactly the case for paths 4 and 5,
due to numerical ``noise'' in the course of practical NEB calculations. From the end points,
the energy profile departs nicely parabolically, which is not so obvious with 13 images but
much better represented for 25-images path (cf. bottom right panel in Fig.~\ref{fig:NEB_results}).
Assuming that the hydrogen atom vibrates \textit{alone} around its
equilibrium position at the extremity of each path, the corresponding frequency 
from fitting the corresponding force constant falls between 776~cm$^{-1}$ (along the path 4 or 5)
to 816~cm$^{-1}$ (along the path 3). 
This nicely matches the prediction from the ``full'' phonon calculation addressed above.

We turn now to a discussion of different paths. 
The paths 4 and 5 seem degenerate; that follows indeed from an observation that
the mid-point in each of them is the same, -- namely, a Ti vacancy site, from which the path
can equally turn ``downwards'' or ``upwards'', consistently to drawings in the two last
panels of Fig.~\ref{fig:path}. One sees a shallow local minimum on top of the barrier, 
already mentioned in subsection~\ref{subsec:energies} as a manifestation of magnetic solution. 
Therefore, this is a metastable position 
for a hydrogen atom, rather than a genuine saddle point. However, there is a subtlety.
If the mid-point of path 5 is placed at the Ti vacancy site by the crystal symmetry, 
the mid-point of path 4 is subject to technical ``drift'', depending on the spring constant 
along the NEB, or other details of the algorithm used. As a result, the path may tend to get 
shortened by displacing its midpoint slightly ``downwards'' (closer to the oxygen atom, 
as is suggested by a drawing of path 4 in Fig.~\ref{fig:path}),
correspondingly climbing the midpoint energy upwards. Such a ``perturbation'' would eventually
destroy the local minimum on top of the barrier. This was not the case in our calculation
which preserved the similarity of the energy profiles along path 4 and path 5.
We note that the barrier height in corresponding NEB calculations, 3.71~eV, is expectedly
close to the difference between previously discussed adsorption energies at the Ti vacancy
and the O vacancy sites, i.e., $+0.75-(-2.87)=3.62\,{\rm eV}$. The mismatch of $0.09\,{\rm eV}$
may be attributed to the technical difference between the constraints imposed in the course
of ``conventional'' conjugate-gradient search and in a NEB calculation.
We emphasize that the profile of path 5 is an ``improved version'' of the energy profile
preliminary scanned in a row of {\sc Siesta} calculations and depicted in Fig.~\ref{fig:path5}.

\begin{figure}[b!]
\centerline{\includegraphics[width=0.48\textwidth]{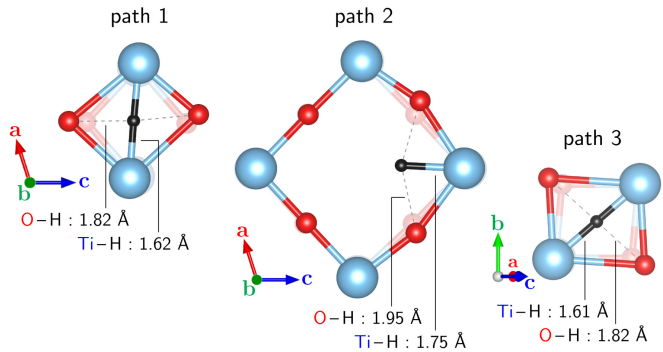}}
\smallskip
\caption{\label{fig:NEB_squeeze}
Optimized structures at the saddle points of paths 1,2, and 3. Distances between
the hydrogen atom and its nearest Ti and O neighbors are indicated. The corresponding
equilibrium structures in the absence of hydrogen are shown faded in the background.
}
\end{figure}

The ``bottleneck'' of the paths 1 and 3 is the hydrogen atom squeezing through the
intact Ti--O--Ti--O square face. The difference is that on the ``upward'' (see Fig.~\ref{fig:path})
path 1 this face is within a $(010)$ lattice plane, whereas for path 3
it makes a $(20\bar{1})$ plane. In path 2, the hydrogen atom passes through an incomplete square face,
missing a Ti atom at one of its corners. The snapshots of the atomic relaxation within
the corresponding planes is shown in Fig.~\ref{fig:NEB_squeeze}, in comparison with the unperturbed
(no hydrogen) situation. One notes that, on passing through the ``bottleneck'', the hydrogen atom
only slightly repels its Ti neighbors (Ti--Ti diagonal increases by ${\sim}3\%$) but
quite considerably (${\sim}27\%$ of the initial distance) pushes apart the O neighbors.
This repulsion can be understood from purely electrostatic arguments, since hydrogen, like oxygen,
is more electronegative than titanium and receives some electron density from the latter.
For this reason, skirting an oxygen atom in the mid-point of path 4 does not bring
this path considerably to the side (downwards in Fig.~\ref{fig:path}), as discussed above, 
whereas the path 2, laid in fact across a cavity with missing Ti atom, comes quite close 
to the remaining Ti atom, which the path skirts.

With the ``bottlenecks'' on path 1 and path 3 being so similar (cf. Fig.~\ref{fig:NEB_squeeze}),
we cannot suggest any obvious reason for the difference in corresponding barrier heights
other than the ``natural'' anisotropy of the crystal structure. In any case, the path 2
undoubtedly possesses the lowest energy barrier, since it does not include a passage
between closely placed two Ti and two O atoms, pushing those latter to the sides. 
Consequently, the path 2 is expected to dominate among hypothetical channels of 
hydrogen diffusion, to which the other paths, possessing the barrier energy of the same order
of magnitude, should contribute in parallel without being a priori excluded.

The height of the barrier measured from the endpoint minima, i.e. the activation energy $E_a$,
may in principle serve to estimate the reaction (e.g., diffusion) rate $D$ via the Arrhenius equation
(see, e.g., the review by Gomer \cite{RepProgPhys53-917} for detail):
\begin{equation}
D=\nu\exp\!\left(\!-\frac{E_a}{k_{\rm B}T}\right)\,.
\end{equation}
A more detailed analysis of different contributions to the flux of hydrogen atoms
induced by gradient of concentrations and involving hoppings over the barriers
was given by Kirchheim and Pundt \cite{PhysMetall25-2597}, culminating in Eq.~(55)
of their work. 
The frequency prefactor $\nu$ (called ``attempt frequency'' in Ref.~\citenum{PhysMetall25-2597})
may vary in very broad ranges, depending
on the reaction type; estimations for proton diffusion in TiO might be not obvious. 
Zhdanov \cite{SurfSciReports12-185} summarized
a number of parameters from the literature in Tables~1 and 2 of his review work;
for instance, for desorption of H$_2$ from Pt $\nu$ makes $10^6\;{\rm s}^{-1}$
and from other different metal surfaces -- $10^{12}$ to $10^{17}\;{\rm s}^{-1}$.
In any case,
huge heights of the barriers relative to the ambient thermal energy
($k_{\rm B}T=26\,{\rm meV}$ at $25^{\circ}$C) in practical sense identifies the diffusion
over barriers as an utterly improbable event. 
A general understanding is that quantum tunneling plays an important role in hydrogen transport 
in metals \cite{PRL37-25,Kehr_1978}. A more detailed analysis along this line would go
beyond the scope of the present work, but it is likely to be responsible for much higher
mobility of hydrogen in different materials than it could be understood in terms of hopping
over the barriers. Coming back specifically to TiO,
Skripov \textit{et al.} \cite{JAlloysComp887-161353} set ${\sim}10^5\,{\rm s}^{-1}$
as the upper limit on the hydrogen jump rate in strongly substoichiometric and
nearly stoichiometric H-doped TiO, referring to the NMR line width: 
``...H atoms in titanium monoxides appear to be immobile on the NMR frequency scale 
up to 370~K''. As an explanation, Skripov \textit{et al.} evoke spatial separation between
the oxygen vacancy sites. We reinforce this conclusion by identifying barrier heights 
as unusually high, by the standards of hydrogen diffusion in metals. 
For comparison, Pozzo et Alf{\`e} \cite{IntJHydrEnergy34-1922} calculated 
the diffusion barriers for hydrogen atoms over the Mg(0001) surface doped with
different transition metals to span the values from nearly zero (Ag doping) to --
the largest among the systems probed -- 0.94~eV (Zr doping) and 0.75~eV (Ti doping).

In view of high barriers separating the hydrogen adsorption sites, and in agreement with
the experimental evidence so far available, TiO seems to be promising for accumulating
hydrogen, even if not so in terms of easiness of the hydrogen diffusion. This can give rise
to interesting applications, with a perspective of extension over other related materials.
Interestingly, the group of Lefort \textit{et al.} reported 
\cite{ElectrochemComm11-2044,IntJHydrEnergy40-8562}
an enormous capacity to accommodate hydrogen (up to 2.9 wt.\%) 
for highly substoichiometric titanium carbide, TiC$_{0.6}$,
which was not the case in weakly substoichiometric TiC$_{0.9}$. The authors attribute 
this property to the presence of long-range-ordered carbon vacancies in TiC$_{0.6}$,
in contrast to TiC$_{0.9}$. In view of the similarity of the crystal structures of titanium
carbide and monoxide, this discovery is stimulating for the extension of our present study.

\section{Conclusions}
\label{sec:conclu}
Summarizing, we performed
first-principles calculations of electronic structure, lattice vibrations
and possible diffusion barriers in monoclinic TiO doped with H atom. 
Our results reinforce the earlier available experimental evidence that
hydrogen atoms enter the oxygen vacancy sites, the related energy gain
now being estimated as ${\simeq}\,2.87\,$eV. Moreover, the Ti vacancy sites 
were identified as possible metastable positions for adsorbed hydrogen atoms, 
unfavorable in energy (with respect to the case of desorbed 
hydrogen) by 0.75~eV. The Ti vacancy site makes a mid-point of two possible diffusion paths
(those with the highest barrier, 3.71~eV) connecting two adjacent O vacancy sites. 
The lowest-energy path (with the barrier height of 2.87~eV) goes around a Ti atom
at a distance of 1.75~{\AA} from it. Different paths involve squeezing the H atom through 
different crystal structure bottlenecks, whereby the O atoms are considerably
pushed away and the Ti atoms are somehow attracted towards the hydrogen atom.
The estimated values of barrier height are too high to account for an appreciable
hydrogen diffusion rate, assuming the hoppings over barriers as the principal diffusion mechanism.
It seems plausible that quantum tunneling processes may play an important role,
as is the case with hydrogen diffusion in other materials.
Calculations of lattice vibration spectrum are consistent with earlier reported results
of inelastic neutron diffusion; the remaining deviations offer a substance for
discussion about the placement of hydrogen atoms within the oxygen-vacancy cages.

Under an angle of possible applications, our study demonstrates that (and explains why)
the Ti monoxide may absorb considerable amount of hydrogen, which however tends to remain immobile
in the sense of diffusivity through the lattice. This might be promising, e.g.,
for superconductivity, the tendency for which can be addressed in a separate study.
As another prospective extension towards
practical needs, the uptake and dissociation of molecular hydrogen at the surface,
an issue almost routinely simulated with some other materials, may seem interesting here. 
In the context of fundamental science, the manifestation of quantum effects in the uptake and diffusion
of hydrogen may deserve a thorough study; moreover, an interplay of vacancy ordering
and diffusion may happen to be interesting.
In any case, an enrichment of experimental evidence will be highly motivating.

\begin{acknowledgements}
The authors thank Dr. Alexander Skripov for careful reading of the manuscript
and for useful comments. The authors thank Elsevier for granting a permission
to reproduce Fig.~5 from the article ``Hydrogen in nonstoichiometric cubic titanium 
monoxides: X-ray and neutron diffraction, neutron vibrational spectroscopy and NMR studies''
by A.V.~Skripov \textit{et al.}, originally published in Journal of Alloys and Compounds
(Copyright Elsevier 2021) Vol.~\textbf{887}, 161353.
S.V.H. and M.R.M. acknowledge a partial financial support by the Research Council 
of the University of Tehran.
S.V.H. and A.P. thank the mesocenter of calculation EXPLOR at the Universit\'e de Lorraine 
(project 2019CPMXX0918) for granting access to computational resources.
\end{acknowledgements}


%

\end{document}